\documentclass[superscriptaddress,preprint,showpacs,preprintnumbers,amsmath,amssymb]{revtex4}

\usepackage{graphicx}
\usepackage{bm}

\begin{document}

\title{Josephson-plasma-resonance and phonon anomalies in trilayer
$\rm Bi_2Sr_2Ca_2Cu_3O_{10}$.}

\date{\today}
\author{A.V. Boris}
\altaffiliation[Also at ]{Institute of Solid State Physics,
Russian Academy of Sciences, Chernogolovka, Moscow distr., 142432 Russia}
\affiliation{Max-Planck-Institut f\"{u}r Festk\"{o}rperforschung,
Heisenbergstrasse 1, D-70569 Stuttgart, Germany}  
\author{D. Munzar}
\affiliation{Institute of Condensed Matter Physics,
Masaryk University, Kotl\'{a}\v{r}sk\'{a} 2, CZ-61137
Brno, Czech Republic}
\author{N.N. Kovaleva}
\altaffiliation[Also at ]{Institute of Solid State Physics,
Russian Academy of Sciences, Chernogolovka, Moscow distr., 142432 Russia}
\affiliation{Max-Planck-Institut f\"{u}r Festk\"{o}rperforschung,
Heisenbergstrasse 1, D-70569 Stuttgart, Germany}
\altaffiliation[Also at]{Institute of Solid State Physics,
Russian Academy of Sciences, Chernogolovka, Moscow distr., 142432 Russia}
\author{B. Liang}
\affiliation{Max-Planck-Institut f\"{u}r Festk\"{o}rperforschung,
Heisenbergstrasse 1, D-70569 Stuttgart, Germany}
\author{C.T. Lin}
\affiliation{Max-Planck-Institut f\"{u}r Festk\"{o}rperforschung,
Heisenbergstrasse 1, D-70569 Stuttgart, Germany}
\author{A. Dubroka}
\affiliation{Institute of Condensed Matter Physics,
Masaryk University, Kotl\'{a}\v{r}sk\'{a} 2, CZ-61137
Brno, Czech Republic}
\author{A.V. Pimenov}
\affiliation{Max-Planck-Institut f\"{u}r Festk\"{o}rperforschung,
Heisenbergstrasse 1, D-70569 Stuttgart, Germany}
\author{T. Holden}
\affiliation{Max-Planck-Institut f\"{u}r Festk\"{o}rperforschung,
Heisenbergstrasse 1, D-70569 Stuttgart, Germany}
\author{B.Keimer} 
\affiliation{Max-Planck-Institut f\"{u}r Festk\"{o}rperforschung,
Heisenbergstrasse 1, D-70569 Stuttgart, Germany}
\author{Y.-L. Mathis}
\affiliation{Forschungszentrum Karlsruhe, Postfach 3640, D-76021 Karlsruhe,
Germany}
\author{C. Bernhard}
\affiliation{Max-Planck-Institut f\"{u}r Festk\"{o}rperforschung,
Heisenbergstrasse 1, D-70569 Stuttgart, Germany}

\date{\today}

\begin{abstract}
The far-infrared (FIR) $c$ axis conductivity of a Bi2223 crystal 
has been measured by ellipsometry. Below T$_c$ a strong absorption band develops 
near 500 cm $^{-1}$, corresponding to a transverse Josephson-plasmon. 
The related increase in FIR spectral weight leads to a 
giant violation of the 
Ferrell-Glover-Tinkham sum-rule. The gain in $c$ axis kinetic energy 
accounts for a sizeable part of the condensation energy.
We also observe phonon anomalies 
which suggest that the Josephson-currents lead to a drastic variation of the
local electric field within the block of closely spaced CuO$_2$ planes.
\end{abstract}
  
\pacs{74.72.Hs,74.25.-q,74.25.Kc,74.50.+r} 

\maketitle
The transition from a normal metal to a superconductor (SC) below the critical
temperature, $T_c$, is accompanied by a redistribution of spectral weight,
($SW$) from finite frequencies in the normal state (NS) into a $\delta$-function at zero frequency in the SC state
that represents the loss-free response of the
SC condensate. For classical SC's the energy gap determines the frequency range
over which the $SW$ of the $\delta$-function is collected, 
so that noticeable changes occur only for $\omega \lesssim 4 \Delta$,
(the so-called Ferrell-Glover-Tinkham (FGT)
sum rule) \cite{Tinkham}. Recently, it was claimed that
the FGT sum-rule is partially violated for the $c$ axis response of the cuprate high-$T_c$
(HTSC) compounds $\rm Tl_2Ba_2CuO_{6+x}$ (Tl2201), $\rm La_{2-x}Sr_xCuO_4$
(LaSr214) and $\rm YBa_2Cu_3O_{6.6}$ (Y123) \cite{Basov}. It was found that the $SW$ loss in
the FIR below $T_c$ is smaller than the $SW$ of the $\delta$-function at
zero frequency, which is independently determined from the imaginary part of the conductivity.
However, the change of the 
FIR-SW in the SC state is small and hard to measure experimentally.
Nevertheless, due to its important implications, this report has attracted considerable attention.
It implies that a very large frequency scale is involved in the SC pairing
and seems to rule out any conventional mechanism that relies exclusivley
on low-frequency bosons like phonons.
Instead, it supports models where an increase in the $c$ axis kinetic energy, $\Delta H_c$,
below $T_c$ provides a significant contribution to the SC condensation energy \cite{Anderson,Chakravarty}.

These far reaching implications call for furher experiments on a compound
where the related $SW$ transfer is larger and therefore more easily identified. 
The most promising candidates are multilayer HTSC compounds, which contain more
than one CuO$_2$ plane per unit cell. 
For the bilayer systems Y123 and
Bi2212 it has already been shown that a sizeable absorption peak develops below
$T_c$ in the FIR range. Evidence has been presented that its $SW$ is mostly electronic in origin and that 
it is part of the SC condensate \cite{Munzar1,Gruninger,Bernhard1,Zelezny,Munzar2}.
This can be understood in terms of the interlayer-tunneling (ILT) model \cite{Anderson}
which assumes that the CuO$_2$ planes
are weakly coupled by the Josephson-currents in the SC state. For bilayer
compounds this results in two kinds of Josephson-junctions with different longitudinal
plasma frequencies \cite{Marel}. Their out-of-phase oscillation gives rise
to a transverse
Josephson-plasma resonance (t-JPR) which has been assigned to the absorption peak that develops
below $T_c$.
This model, termed Josephson-superlattice-model (JSL) in the following,
has been successfully applied to describe the anomalous FIR $c$
axis response of Y123 and $\rm Bi_2Sr_2CaCu_2O_{8}$
(Bi2212) \cite{Munzar1,Gruninger,Bernhard1,Zelezny,Munzar2}. Nevertheless,
it is not commonly accepted yet, and it is even disputed whether the $SW$ of the absorption band
is electronic in origin, and whether it arises from
higher frequencies \cite{Homes}. 

In this manuscript we present ellipsometric data of the $c$ axis dielectric response 
of the trilayer compound $\rm Bi_2Sr_2Ca_2Cu_3O_{10}$ (Bi2223). We show
that a strong absorption band develops below $T_c$ in the FIR, corresponding
to a t-JPR. It leads to a clear increase of the $SW$ in the FIR below $T_c$
corresponding to a giant violation of the FGT sum-rule.
We quantify the associated change in kinetic
energy and show that it can account for a substantial part of the SC condensation energy.
We also show that the electronic mode and the associated phonon anomalies can be qualitatively described
with the JSM. The drastically different SC induced
anomalies of the two similar oxygen
bond-bending modes provide clear evidence that the local electric
field can vary significantly along the $c$ axis direction even within a stack
of closely spaced CuO$_2$ layers and thus provide strong support for the
JSM. 

A Bi2223 single crystal of dimensions 6 $\times $ 4 $\times $0.5 mm$^3$
was grown by the travelling solvent floating zone technique.
The crystal contains more than 95 \% of Bi2223 with only a minor fraction of layer-intercalated
Bi2212 \cite{Liang}.
In its as grown state, the crystal was underdoped with $T_c$ = 97 K (midpoint)
and $\Delta T_c$ = 7 K. Subsequent to the optical measurements
the same crystal was annealed for ten days in flowing oxygen at 500$^\circ $C
(and then for three days in air at 700$^\circ $C)
and rapidly quenched so 
it was nearly optimally doped with $T_c$ = 107  K and $\Delta T_c$ = 3 K
(moderately underdoped with $T_c$ = 102 K and $\Delta T_c$ = 4 K).
The ellipsometric measurements
(see Ref. \cite {Bernhard1} for a description of the technique)       
have been performed at the infrared beamline
of the synchrotron radiation source at ANKA in Karlsruhe, Germany and at
NSLS in Brookhaven, USA.
A home-built ellipsometer attached to a "Bruker" IFS
66v/S FT-IR spectrometer has been used.
The high brilliance of the synchrotron enables us
to obtain accurate ellipsometric data in the FIR spectral range
even on mm-sized samples.

Figure 1 shows the real part $\sigma_{1}(\omega)$ of the $c$ axis optical conductivity
of Bi2223 at the three different doping levels.
Shown are spectra for one temperature somewhat above $T_c$ and for several temperatures below $T_c$.
The normal-state spectra are dominated by the contributions of several IR-active phonons,
the one of the charge carriers is very weak. The main phonon bands are located
at 97, 128 \cite{notshown}, 170, 211, 305, 360, 400 and 582 cm$^{-1}$.
Except for the additional modes of Bi2223 at $400 {\rm\,cm^{-1}}$ and $128 {\rm\,cm^{-1}}$
IR-active phonons appear in Bi2212 at similar frequencies \cite{Zelezny,Munzar2}.
In order to find the corresponding eigenvector patterns we have performed shell
model calculations for the Bi-based compounds in the body centered tetragonal $I4/mmm$ structure.
We obtained a set of parameters which allows to reproduce
simultaneously the structures, the experimental values of the $c$ axis dielectric constants
and the frequencies of the IR-active modes in Bi2223, Bi2212 and Bi2201.
For Bi2212 our assignment of the modes agrees well with the one given in Ref. \cite{Prade}.
Details will be presented elsewhere \cite{Kovaleva}.
The strongest modes at 582 and 305 cm$^{-1}$ involve primarily $c$ axis vibrations
of the apical O2 oxygen and the Bi-plane O3 oxygen, respectively.
The eigenvector patterns of the modes at 360 and 400 cm$^{-1}$
corresponding to the in-phase and the out-of-phase motion
of the oxygens in the inner and the outer $\rm CuO_2$ layers
are shown in Fig. 3(a).
The remaining four $A_{2u}$ IR-active phonon modes at lower frequencies
involve vibrations of the heavy ions and will not be further discussed here.
Some additional weak modes at 276, 471 and 635 cm$^{-1}$
are most likely related to 
the incommensurate modulation in the BiO and SrO layers\cite{Jacubowicz}.

In the NS the spectra exhibit hardly any noticeable changes, except
for a sharpening of the phonons with decreasing temperature (not shown).
Right below $T_c$, however, the spectra change appreciably.
This is also illustrated in Fig. 2(a) which displays
the difference
$\sigma_1(T = 10 K) - \sigma_1(T = 120 K)$.
The most prominent feature is
the broad absorption band around $500{\rm\,cm^{-1}}$
which appears below $T_{c}$ and grows rapidly with decreasing temperature.
Figures 1 and 2(a) show that the center of this band shifts towards higher
frequencies with increasing doping.
A similar absorption band has been previously identified
in the bilayer compound Y123 where it has been assigned \cite{Munzar1,Gruninger,Bernhard1}
to the t-JPR of the JSM. While this feature 
is rather weak in Bi2212 \cite{Zelezny,Munzar2}, its $SW$
is very large in Bi2223 and gives rise to a considerable increase in the FIR-$SW$ below T$_c$.
This is evident from Fig. 2(b) where we show the difference
between the $SW$ in the SC state at 10 K and the one in the NS at 120 K:
\begin{eqnarray}
\Delta SW(\omega) =
\int_{100\,cm^{-1}}^{\omega}{(\sigma_1(10,\omega')-\sigma_1(120,\omega'))d\omega^{'}}\,
\label{eq:one}
\end{eqnarray}
Apart from some smaller changes related to the phonon anomalies (as discussed below)
it exhibits a steep increase between $400{\rm\,cm^{-1}}$ and $550{\rm\,cm^{-1}}$.
Above 650 cm$^{-1}$ $\Delta SW(\omega)$
remains essentially constant (as shown in Fig.2 up to 1300 cm $^{-1}$ for
the optimally doped sample) at $\Delta SW$ = 800, 1100, and 1400 $\Omega^{-1}{\rm cm^{-2}}$
for $T_{c}$ of 97, 102, and 107 K, respectively.
Such an apparent increase in the FIR-$SW$ below $T_{c}$
is certainly not expected for any conventional SC
where FIR-$SW$ should be removed and transferred
to the $\delta$-function at zero frequency with $(N_n-N_s)/\rho_s=1$ (FGT sum-rule) \cite{Tinkham},
where $N_n-N_s =-120/\pi \cdot \Delta SW(\Omega_c)$ (cutoff frequency
$\Omega_{c}\gtrsim 4\Delta$), $\rho_s$ is the SC condensate density.
For Bi2223 upper limit of $\rho_s $ can be estimated using 
the London penetration depth $\lambda_c=23 ~ \mu m$ for $\rm Bi_{1.85}Pb_{0.35}Sr_2Ca_2Cu_{3.1}O_y$
\cite{Shibata}: $\rho_s =(c/\lambda_c)^2\lesssim 4800\rm ~ cm^{-2}$.
The inset of Fig.2(b) shows that our data on Bi2223 with $T_c$ = 107 K in terms
of the FGT sum-rule give $(N_n-N_s)/\rho_s\lesssim-10$ at $\Omega_{c}\gtrsim 650~
\rm cm^{-1}$. The data certainly represent a striking manifestation of the violation of the FGT sum rule.
They highlight that a significant amount of $SW$ is transferred from higher frequencies
to the t-JPR in the FIR. We emphasize that the $SW$ of the t-JPR
is part of the SC condensate just as much as the $\delta$-function at zero frequency.
For Bi2223 the $SW$ of the t-JPR, however, exceeds the one of the $\delta$-function at zero frequency
by one order of magnitude.

It is evident from Fig. 1 and Fig. 2(a) that the formation of the t-JPR
is also associated with an anomalous temperature dependence of some of the phonon modes,
in particular, of the modes at 360, 400 and 582$ {\rm\,cm^{-1}}$ denoted
by $A,B,C$ in Fig. 2(a). Particularly interesting are the contrasting $T$ dependences of the in-phase
and out-of-phase oxygen bond-bending modes at 360 and 400${\rm\,cm^{-1}}$ 
whose eigenvector diagrams are shown in Fig. 3(a).
As shown in Fig. 3(b), the mode at 360 cm$^{-1}$ loses a significant amount 
of its $SW$ below $T_{c}$. 
A similar effect has been observed for the oxygen bond-bending mode in Bi2212.
In clear contrast, the out-of-phase mode at 400 cm$^{-1}$ (which is specific to the trilayer compound)
gains a significant amount of $SW$ in the SC state. In the following we show that
this behavior, while surprising at first, is explained by the JSM.

As outlined in Refs.\cite{Munzar1,Bernhard1,Zelezny,Munzar2}
the onset of the Josephson-currents between the CuO$_{2}$ layers can
lead (in the absence of a significant screening in the NS)
to a significant change of the 
dynamical local electric field.
A simple estimate of the local field can be obtained by considering a stack
of homogeneously charged Josephson coupled CuO$_{2}$ layers. A sketch is
shown in Fig. 3(c) where $\kappa (\omega)$ denotes the charge density that
alternates from one outer plane to the other. 
The Josephson-currents $j_{bl}(\omega)$ and $j_{int}(\omega)$ can be described by using the
local dielectric functions of the intra-trilayer region and of the inter-trilayer region,
$\varepsilon_{tl}(\omega)=\varepsilon_{\infty}-\omega_{bl}^{2}/\omega^{2}$ and
$\varepsilon_{int}(\omega)=\varepsilon_{\infty}-\omega_{int}^{2}/\omega^{2}$,
respectively. Following the model of Ref. \cite{Munzar1} the normalized
local field inside the spacing layer that separates the trilayers, $E^{\star}_{int}$,
inside the trilayer, $E^{\star}_{in}$, and at the outer CuO$_{2}$ layers, $E^{\star}_{out}$
are  :
$$
E^{\star}_{int} = {E_{int}\over <E>}={(d_{tl}+d_{int})\varepsilon_{tl}\over
(d_{tl}\varepsilon_{int}+d_{int}\varepsilon_{tl})},
\eqno (3a)
$$
$$
E^{\star}_{in} = {E_{in}\over <E>}={(d_{tl}+d_{int})\varepsilon_{int}\over
(d_{tl}\varepsilon_{int}+d_{int}\varepsilon_{tl})},
\eqno (3b)
$$
$$
E^{\star}_{out} = {(E^{\star}_{int}+E^{\star}_{in})\over 2},
\eqno (3c)
$$
where $<E>$ is the average field, $d_{tl}=2d_{bl}$.
For $\omega_{int}=0$, $\omega_{bl}=1250{\rm\,cm^{-1}}$,
$d_{int}=12{\rm\,\AA}$, and $d_{bl}=3.37{\rm\, \AA}$
we obtain the result that is shown in Fig. 3(d).
In the NS with $\omega_{int}=0$ and $\omega_{bl}=0$
we have $E^{\star}_{int}=E^{\star}_{in}=E^{\star}_{out}=1$. Note the following:
(i) In the frequency range of the bond-bending modes (360 to 400 cm$^{-1}$) $E_{out}$
is positive but strongly suppressed with respect to the NS, whereas $E_{in}$
changes sign at $T_c$ and acquires a large negative value;
(ii) The value of $E_{int}$ around the frequency of the apical mode (582 cm$^{-1}$)
is also strongly suppressed with respect to the NS.

The $SW$ of a given phonon mode is determined by the local field at the ions
participating in the mode and by the mode polarizability. 
The trends (i) account for the anomalies of the bending modes. 
Concerning the in-phase mode, the pattern of the local field is in agreement
with the eigenvector pattern above $T_c$ ($E_{out}$ and $E_{in}$ are parallel) but not below $T_c$
($E_{out}$ and $E_{in}$ are antiparallel). In addition, the magnitude of $E_{out}$ is
strongly suppressed below $T_c$. Both effects lead to the observed decrease of
the $SW$ of the in-phase mode in the SC state. For the out-of-phase mode the situation is reversed,
that is, the two patterns are not in agreement above $T_c$ whereas they are in accord below $T_c$. In addition,
the magnitude of $E_{in}$ increases below $T_{c}$. Both effects contribute to the $SW$
increase of the mode below $T_c$. We emphasize that these results are rather robust 
against a change of the relative amplitudes of the ionic displacements
which will only affect the relative contribution of the above mentioned
effects. 
Finally, point (ii) allows one to understand
why the $SW$ of the apical oxygen mode at $582{\rm\,cm^{-1}}$
decreases below $T_c$: $E_{int}$  
decreases below $T_{c}$ and so does the $SW$.
We can also understand why the anomaly is much stronger
in Bi2223 than in Bi2212.
For $\omega_{int}=0$ and for $\omega$ considerably larger than the frequency
of the t-JPR, $\omega_{pl}=\omega_{bl}\sqrt{d_{int}/[(d_{tl}+d_{int})\varepsilon_{\infty}]}$,
it follows from Eq.~(3a) that 
$E^{\star}_{int}\approx 1-[d_{tl}\omega_{pl}^{2}/(d_{int}\omega^{2})]$. 
The corresponding equation for the bilayer system reads
$E^{\star}_{int}\approx 1-[d_{bl}\omega_{pl}^{2}/(d_{int}\omega^{2})]$. 
The magnitude of the second term on the right hand side which is responsible 
for the suppression of $E^{\star}_{int}$ at high frequencies 
is twice as high for the trilayer system than for the bilayer one. This
explains the stronger reduction of the $SW$ of the apical oxygen below $T_{c}$
in Bi2223 than in Bi2212. Note that these arguments merely depend on the trilayer geometry 
and on the circumstance  that the frequencies of the bending modes 
are lower than $\omega_{pl}$ whereas the frequency of the apical oxygen mode is somewhat higher. 
These phonon anomalies and the underlying changes of the local electric field
in the SC state clearly reflect a transition from a state exhibiting confinement 
(incoherent intra-trilayer conductivity) into a state where the CuO$_{2}$ planes are Josephson-coupled.   
We demonstrate that in the SC state the local electric field can exhibit enormous variations within the unit cell,
even its sign can change between the inner- and outer- layers of the trilayer. We are not aware
of any other model which would allow one to describe these phenomena in so
much detail.  

Having shown that the JSM provides an excellent description of the electronic
spectra and the phonon anomalies of Bi2223, we now use this model to evaluate
the change in the $c$ axis kinetic energy of the charge carriers, $\Delta H_c$,
which is related to the growth of the t-JPR. 
In a recent paper \cite{Munzar2} 
$\Delta H_c$ in Y123 and Bi2212 has been estimated
according to $\Delta H_c=E_J$, where $E_{J}$ is the coupling energy
of the intra-bilayer Josephson junctions.
This approach, however, ignores the changes upon entering the SC state
due to the single particle tunnelling and thus may overestimate the value of $\Delta H_c$.
The more rigorous sum-rule approach \cite{Chakravarty}
is appropriate only for single-layer materials but not for the bi- or trilayer ones
\cite{Munzar2}.
In the meantime some of us have derived \cite{Munzar3}
a version of the sum-rule that is valid for multilayer compounds
with fully insulating blocking layers such as Bi2212 or Bi2223.
For trilayer cuprates it reads
$$
\Delta H_{c}={\hbar^2\varepsilon _0 a^2\over  e^2 }
{(2d_{bl}+d_{int})\over d^2_{bl}}
\Delta SW(\Omega _c),
\eqno (2)
$$
where $d_{bl}(2223)=d_{bl}(2212)$, 
$a$ is the in-plane lattice constant.
Using the results shown in Fig. 2(b) we obtain
$\Delta H_{c}\approx $ 0.06, 0.08, and 0.11 meV for
$T_{c}=$ 97, 102, and 107 K respectively. The condensation energy of Bi2223 
is not known. Remarkably, however, our results for $\Delta H_{c}$ are
comparable to the condensation energy of $0.13{\rm\,meV}$ obtained by
specific heat measurements for optimally doped Bi2212 \cite{Loram}.

In summary, our data of the $c$ axis dielectric function of Bi2223
provide clear evidence that the t-JPR is a universal feature 
of the multilayer HTSC cuprate compounds.
They show  unambiguously that the $SW$ of the t-JPR is electronic in origin and arises 
from high frequencies beyond the FIR range. We have shown that the related transfer
of $SW$ gives rise to a significant change in $c$ axis kinetic energy of the charge carriers
which can account for a sizable part of the SC condensation energy.
We also observe phonon anomalies which suggest that the Josephson-currents lead to a strong
variation of the dynamical local electric field even
between the inner and outer CuO$_2$ planes of a trilayer.
\begin{acknowledgments}
We thank L.Carr for the support at NSLS.
T.H. acknowladges support by the AvH Foundation.
\end{acknowledgments}

\newpage

\begin{figure}
\includegraphics*[width=120mm]{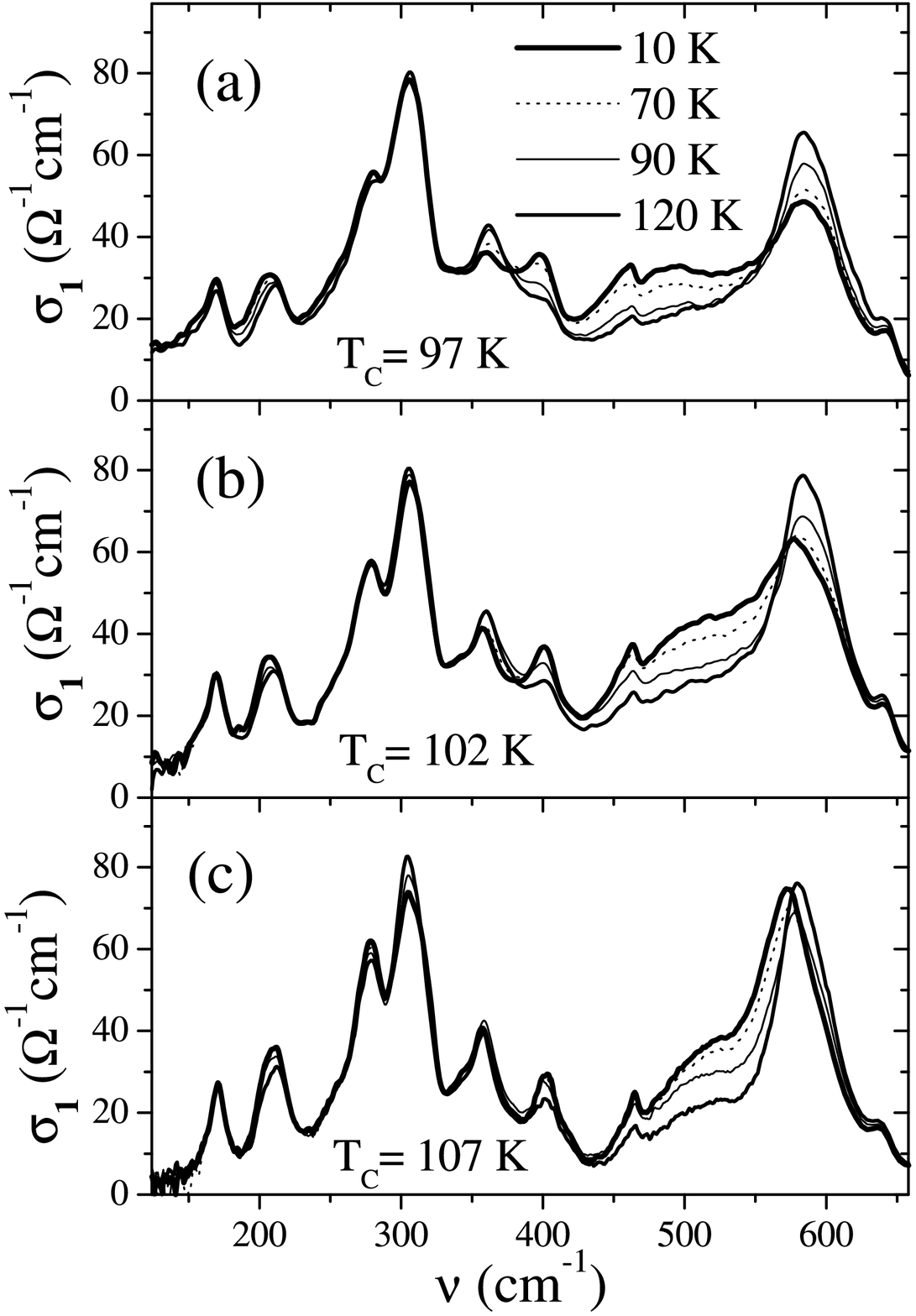}
\caption{Real part, $\sigma_{1}(\omega)$, of the FIR $c$ axis conductivity
of Bi2223.}
\label{Fig1}
\end{figure}

\newpage

\begin{figure}
\includegraphics*[width=120mm]{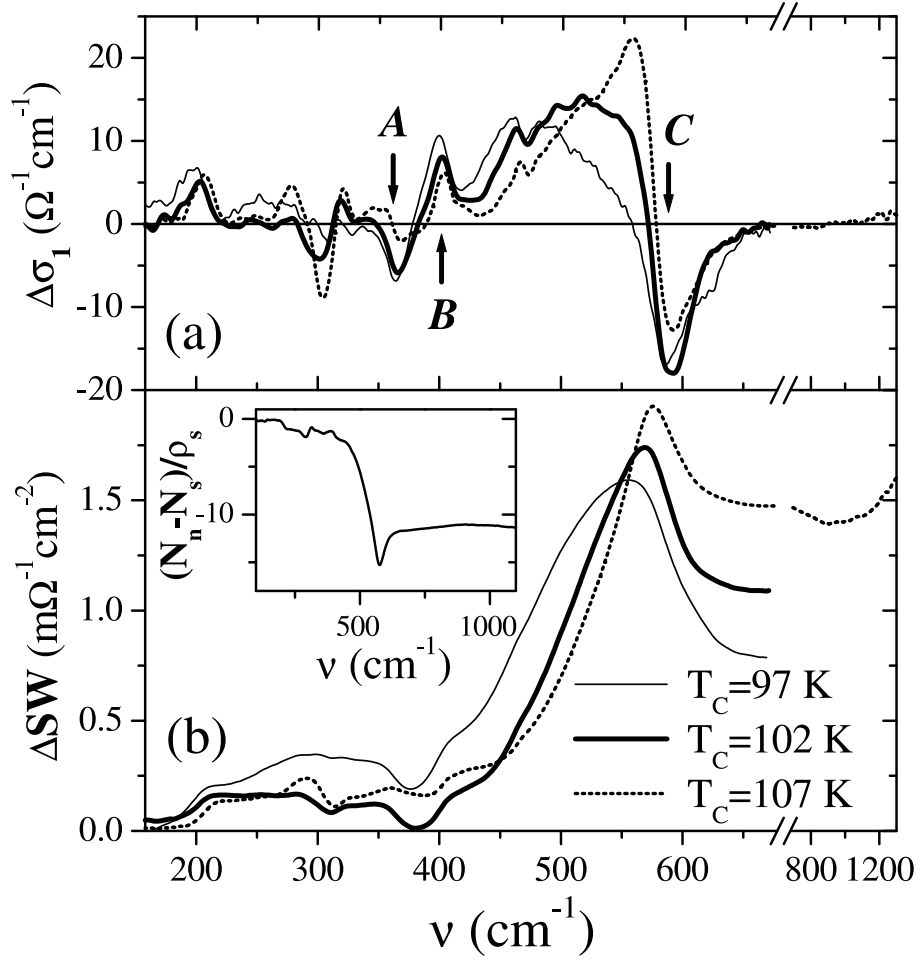}
\caption{Spectra of the differences
(a) $\Delta \sigma_1=\sigma_1 (10\rm\,K,\omega) - \sigma_1(120\rm\,K,\omega)$ and
(b) $\Delta SW(\omega)$ (see Eq.~(\ref{eq:one})).
The phonon anomalies discussed in the text are denoted by {\it A, B} and {\it C}.
Inset: Difference $N_n-N_s =-120/\pi \cdot \Delta SW(\omega )$ for Bi2223 with $T_c$
= 107 K normalized by $\rho_s = 4800~\rm cm^{-2}$.}
\label{Fig2}
\end{figure}

\newpage

\begin{figure}
\includegraphics*[width=120mm]{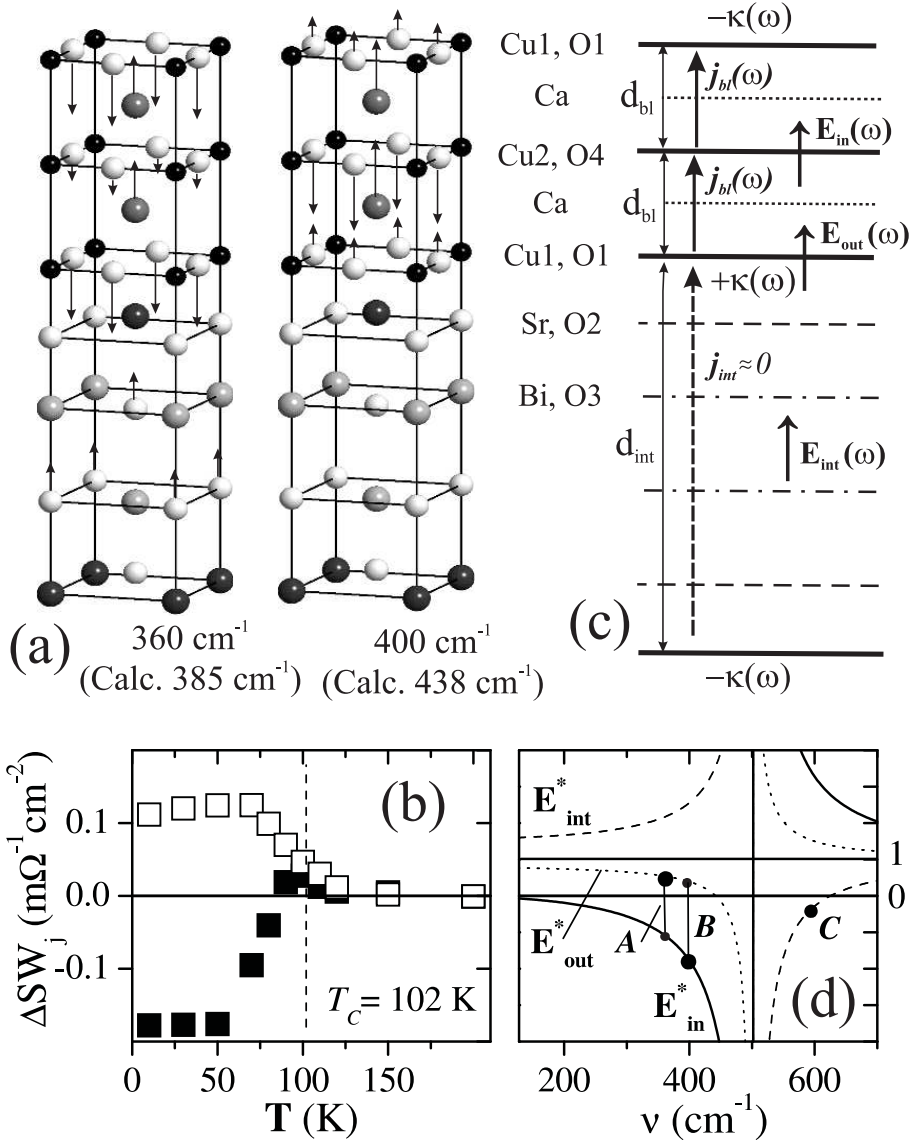}
\caption{
(a) Calculated oxygen bond-bending $A_{2u}$ eigenmodes of Bi2223.
(b) Relative $SW$ changes of the phonons at 360
cm$^{-1}$ (solid squares) and 400 cm$^{-1}$ (open squares) with decreasing
temperature, $\Delta SW_j=(SW_j(T)-SW_j(200K))/SW_j(200K)$.
$SW_j$ has been 
derived by fitting a sum of Lorentzian functions
to the complex dielectric function.
(c) Schematic representation of the model discussed in the text. The horizontal
lines represent CuO$_2$ (solid), Ca (dotted), SrO (dashed), and BiO (dash-dotted) planes.
(d) Frequency dependent variation of the normalized local electric field 
due to the Josephson-currents, $j_{bl}$ and $j_{int}$, as defined in the text:
$E^{\star}_{int}$, $E^{\star}_{in}$ and $E^{\star}_{out}$ are the local
fields inside the intertrilayer spacing (dashed line), inside the trilayer (solid line), 
and at the outer CuO$_{2}$ layers (dotted line), respectively.
The local fields at the site of the particular ions participating in the eigenmodes
at 360, 400 and 582 cm$^{-1}$ are indicated by {\it A, B} and {\it C}, respectively.
}
\label{Fig3}
\end{figure}


\begin{thebibliography}{30}
\bibitem{notshown}Two low-frequency modes are clearly observed in the real part
of the dielectric function $\varepsilon _1 (\omega)$ (not shown).
\end{thebibliography}
\end{document}